\documentclass[aps,prl,reprint,groupedaddress,amsmath,floatfix]{revtex4-2}
\usepackage{amssymb,graphicx}
\usepackage{xcolor}
\newcommand{\ds}{\displaystyle}

\newcommand{\T}{\hat{T}{}}
\newcommand{\ttt}{\hat{t}{}}
\newcommand{\x}{\hat{x}{}}

\begin{document}

%\preprint{}

\title{Temperature Periodic Modulation Doubles the Power Output of Thermoelectric Generators}


\author{Dario Narducci}
\email[]{dario.narducci@unimib.it}
\author{Antonio Mazzacua}
\author{Federico Giulio}
%\thanks{}
\affiliation{University of Milano Bicocca, Dept.\ Materials Science, via R.\ Cozzi 55, I--20125 Milan (Italy)}


\date{\today}

\begin{abstract}
Thermoelectric heat harvesting provides in principle a flexible, reliable way to convert heat into electric energy, thus recovering waste heat and powering off-net devices. However, thermoelectricity conversion efficiency is still low, despite the major efforts deployed over the last two decades to improve materials performances. A largely unexplored opportunity to increase harvesting efficiency is offered by converting heat flux dynamically. It was shown [D. Narducci \textit{et al.}, Mater.\ Today Phys.\ 54, 101713 (2025)] that when the temperature of the heat source is sinusoidally modulated, this leads to an increase of the maximum power by up to 50 \%. In this paper we show how this approach can be further and significantly enhanced. We found that sinusoidal modulation can be outperformed by a different periodic modulation, leading to an increase of the power output by up to 100 \% compared to the stationary case. Furthermore, we extended our analyses beyond the constant-property approximation, showing that the power output enhancement is robust, applying to any real material where the temperature dependence of all phenomenological coefficients is accounted for. This major power output enhancement offers a novel, immediately applicable pathway to overcome the efficiency limits of thermoelectric devices.
\end{abstract}

%\keywords{}

\maketitle

\section{Introduction}

Efficiency of cyclic engines is bound from above by Carnot's efficiency $\eta_\text{C} = 1-T_\text{c}/T_\text{h}$, where $T_\text{h}$ and $T_\text{c}$ ($< T_\text{h}$) are the temperatures of the two thermostats exchanging heat with the working fluid. As known, reaching $\eta_\text{C}$ with real-world engines is impossible but also useless, since Carnot’s engine delivers no power due to the infinite duration of a cycle. Beginning in the sixties, researchers have advanced thermodynamics to cover cyclic heat engines operating at non-zero power, targeting efficiency at maximum power (EMP) \cite{Schmiedl2008,Esposito2010} instead of maximum efficiency. Indeed, heat can be also converted into useful work by non-cyclic engines. Typical examples are photovoltaic cells \cite{DeVos1992,DeVos2008} and thermoelectric generators (TEGs) \cite{Goupil2016book}. They share their capability of generating electric power with no moving parts, therefore with an extended lifetime, and being renewable energy sources. 
Concerning TEGs, their efficiency is ruled by the time-dependent Domenicali's equation \cite{Domenicali1954RMP}:
\begin{equation}
{c_V}\dot{T}(\vec{r},t) = \vec{\nabla}\cdot\left(\kappa_{\rm oc} \vec{\nabla} T(\vec{r},t)\right) +\frac{ |\vec{J}(t)|^2}{\sigma_T}
-\tau \vec{J}(t)\cdot \vec{\nabla}T(\vec{r},t)
\label{eq:tdDE}
\end{equation}
where $T(\vec{r},t)$ is the temperature in $\vec{r}$ at time $t$, ${c_V}$ is the isochoric specific heat per volume unit, $\kappa_{\rm oc}$ is the open-circuit thermal conductivity, $\vec{J}$ is the current density, $\sigma_T$ is the isothermal electric conductivity, $\tau=T\text{d}\alpha/\text{d}T$ and $\alpha$ are Thomson and Seebeck coefficients, respectively. 
For more than 70 years the analysis of power generation was almost entirely developed assuming a constant temperature difference $\Delta T$ across the medium (stationary conditions). 
Neglecting contact resistances and lateral heat dissipation and assuming that $\sigma_T$, $\kappa_{\rm oc}$, and $\alpha$ are all constant in the temperature range across which the TEG operates (constant-property approximation -- CPA), then $\tau\equiv 0$ and the maximum efficiency computes to \cite{NarducciBook2018}
\begin{equation}
\eta_\text{max}=\eta_\text{C}\frac{\sqrt{1+z\bar{T}}-1}{\sqrt{1+z\bar{T}}+T_\text{c}/T_\text{h}}
\end{equation}
where $z=\sigma_T\alpha^2/\kappa_\text{oc}$, $T_\text{h}$ and $T_\text{c}$ are the temperatures at the hot and cold leg ends, and $\bar{T}=(T_\text{h} + T_\text{c})/2$.  The pertinent output power density is 
\begin{equation}
J_\text{w}=\frac{\sigma_T\alpha^2(\Delta T)^2}{2\ell}\frac{\sqrt{1+z\bar{T}}}{1+\frac{1}{2}z\bar{T}+\sqrt{1+z\bar{T}}}
\end{equation}
where $\Delta T=T_\text{h} - T_\text{c}$ is the temperature difference applied across a distance $\ell$. %and $\zeta=2+(1+z\bar{T})^{-1/2}+(1+z\bar{T})^{1/2}$. 
Such maximum efficiency is reached when the ratio $m\equiv R_\text{L}/r$ between the electrical load resistance $R_\text{L}$ and the medium resistance (often referred to as leg) $r$ is $\sqrt{1+z\bar{T}}$. 
Instead, optimizing the output power density leads to the efficiency at maximum power (attained for $m=1$) \cite{NarducciBook2018}:
\begin{equation}
\eta_\text{MP}=\frac{\eta_\text{C}}{2}\left(1+\ds\frac{2}{zT_\text{h}}-\ds\frac{\Delta T}{4T_\text{h}}\right)^{-1}
\end{equation}
at which $J_\text{w}=\sigma_T\alpha^2\Delta T^2/(4\ell)$.

In spite of its seeming simplicity, the analysis of the fundamental physics governing dynamic operation of TEGs remained largely locked behind time-dependent Domenicali's unsolved nonlinear differential equation, a mathematical barrier that defeated every analytical attempt. 
In 1960, Gray \cite{Gray1960} provided an analytical solution to the linearized time-dependent problem, namely in the $|\vec{J}|\to 0$ limit. While observing a phase delay between $\vec{J}$ and the heat flux $\vec{J}_\text{q}$, Gray reported no significant change of the EMP under dynamic conditions. At the opposite, in 2013 Yan and Malen \cite{Yan2013} attempted anew to evaluate the possible advantages of operating a TEG by applying a sinusoidally time-modulated $\Delta T(t)$, also comparing the power output (computed through measurements of the thermovoltage) with a still linearized solution of Domenicali's equation. They reported an increase in the power density close to 80\%, in striking contrast with Gray's results. 
We show here through the exact analytical solution of the time-domain Domenicali's equation how such a dynamic enhancement can double the EMP of any TEG, independently of the thermoelectric materials.  
In a previous paper \cite{Narducci2025} we provided an exact analytical solution of the time-dependent Domenicali's equation when the temperature drop across the thermoelectric element is sinusoidally modulated, reporting an increase of the power output up to 50\% compared to the stationary case. 
We also noted {\it passim} that the choice of a sinusoidal modulation was just the simplest choice, and that alternate periodic modulation functions could have enabled even larger EMP increases.
Here we report dynamic EMP enhancements following the optimization of the modulation function. Exact analytical solution of the time-dependent Domenicali's equation shows that the output power density may be increased by up to 100\% compared to the stationary case. Note that this is equivalent to a threefold increase of the figure of merit $z\bar{T}$ -- thus making available e.g.\ a thermoelectric material converting heat around room temperature with an equivalent $zT$ of up to 3.2. These results will be extended also beyond the CPA, showing their robustness even when the applied temperature difference is large.

\section{Dynamic Efficiency of Thermoelectric Generators \label{sec:anal}}

\begin{figure}
\includegraphics[width=\columnwidth]{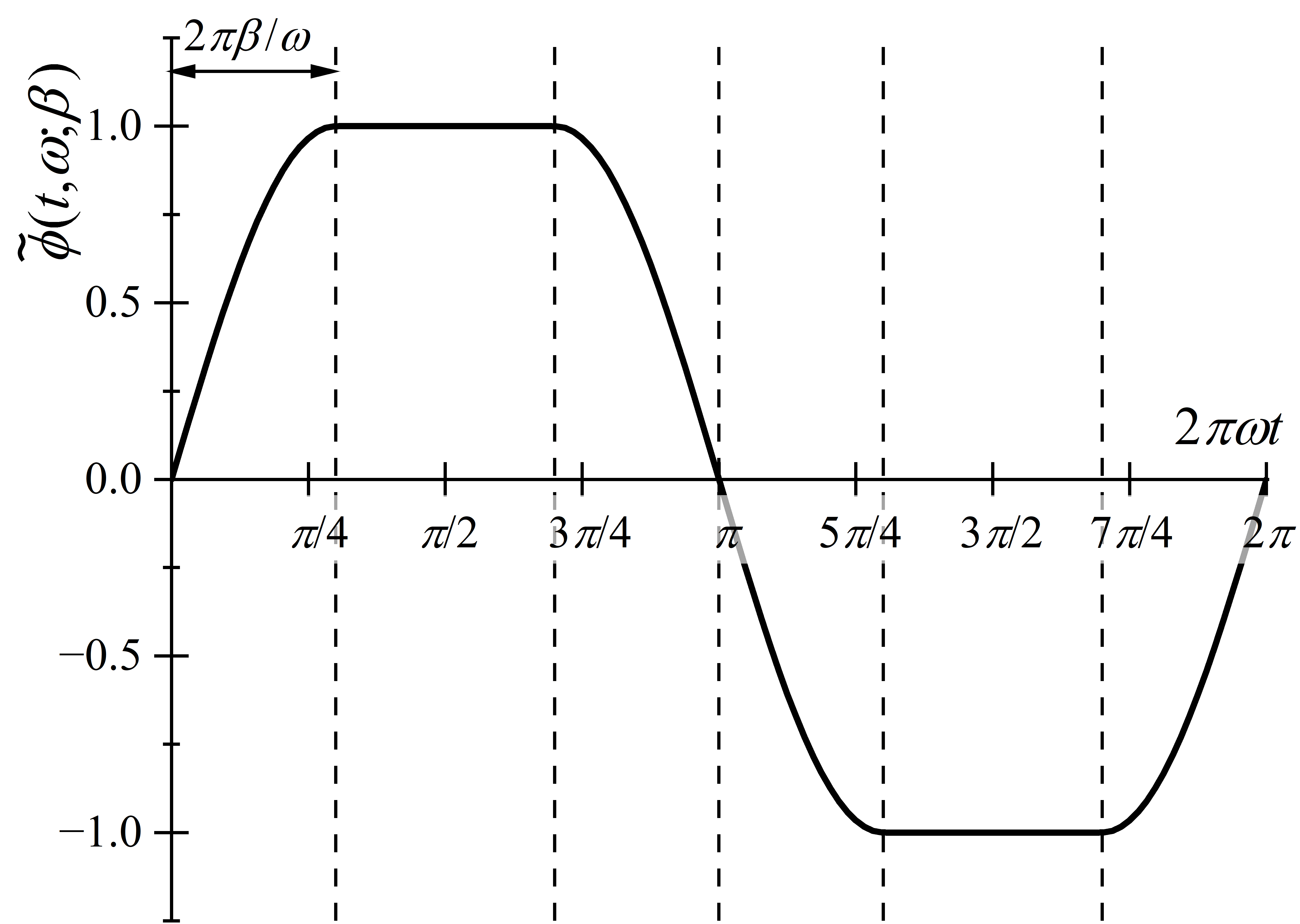}
\caption{Plot of $\tilde{\phi}(t,\omega;\beta)$.}
\label{fig:squarewave}
\end{figure}

Let us  consider the integration of the one-dimensional time-dependent Domenicali's equation (Eq.\ \eqref{eq:tdDE}) under Dirichlet boundary conditions
%\begin{subequations}
\begin{equation}
T(0,t)=T^{(0)}_\text{h} +T^{(1)}_\text{h} \phi(t,\omega;\beta), \qquad  %\label{eq:DBC1}
%\end{equation}
%\begin{equation}
T(\ell, t) = T_\text{c} %\label{eq:DBC2}
\label{eq:DBC12}
\end{equation}
%\end{subequations}
We choose the periodic modulation function $\phi(t,\omega) \in [-1, 1]$ to be zero-averaged and to maximize $\int_0^{2\pi/\omega}\phi(t,\omega)^2 \text{d}t$. A simple variational analysis shows that the optimal $\phi(t,\omega)$ is a square wave function. Thus, we define a parametric function on $t \in [0, 2\pi/\omega]$:
%\begin{widetext}
\begin{equation}
\tilde{\phi}(t,\omega;\beta)=\left\{
\begin{array}{ll}
\sin \left(\ds\frac{\omega t}{4 \beta }\right) 
& 0 \le \varphi \le \beta \\
1 
& \beta < \varphi \le \ds\frac{1}{2}-\beta \\
-\sin \ds\left(\frac{\omega t-\pi}{4 \beta }\right) 
& \ds\frac{1}{2}-\beta < \varphi \le \frac{1}{2}+\beta  \\
-1 
& \ds\frac{1}{2}+\beta < \varphi \le  1-\beta \\
\sin \left(\ds\frac{ \omega t }{4 \beta }\right) 
& 1-\beta < \varphi \le 1
\end{array}
\right.
\end{equation}
where $\varphi=\dfrac{\omega t}{2 \pi}$ and $\beta\in [0,1/4]$. Its domain is extended to $t \in (-\infty, +\infty)$ as $\phi(t,\omega;\beta)= \tilde{\phi}(t-(2 \pi/\omega \lfloor t/(2 \pi)\rfloor)/\omega,\omega;\beta])$. Note that $\phi$ returns a sine function for $\beta=1/4$, degenerating into a square wave function for $\beta=0$ (Fig.\ \ref{fig:squarewave}).
%\end{widetext}

The procedure we follow for the integration of Eq.\ \eqref{eq:tdDE} under boundary conditions \eqref{eq:DBC12} is largely reminiscent of that reported in \cite{Narducci2025}. We intentionally neglect contact thermal and electric  resistances, since the aim is to compare ideal stationary and dynamic thermoelectric heat conversion. It is convenient to introduce the following reduced variables:
\begin{equation} \label{eq:reduced} 
\begin{array}{llll} 
&\hat{J}_{{\rm q}} =J_{{\rm q}} /J_{{\rm q, r}}  &\hat{J}_{{\rm w}} =J_{{\rm w}} /J_{{\rm q, r}}  &J_{{\rm q, r}} =\kappa T_{{\rm h}}^{(0)}/\ell\\ 
&\hat{J}_{{\rm f}} =J_{{\rm f}} /J_{{\rm f, r}}  &J_{{\rm f, r}} =\sigma_T \alpha T_{{\rm h}}^{(0)}/\ell & \\ 
&\hat{T}=T/T_{{\rm h}}^{(0)}  &\hat{x}=x/\ell  &\hat{t}=t/\tau_{{\rm th}} \\ 
&\hat{\omega }=\omega \tau_{{\rm th}} &\tau_{{\rm th}} =\ell ^{2} c_{{V}} /\kappa  & 
%&\zeta =\displaystyle\frac{\sigma_T \alpha^{2} T_{{\rm h}}^{(0)} }{\kappa}  &\beta =\displaystyle\frac{\varepsilon_{0} \kappa }{\ell^{2} c_{{V}} \sigma_T } & 
\end{array} 
\end{equation} 
with $\zeta = \sigma_T \alpha^{2} T_{{\rm h}}^{(0)} /\kappa$. 
Thus, the Dirichlet problem reads
\begin{equation}
\partial_{\hat{t}}{\hat{T}}(\hat{x},\hat{t})-\partial_{\hat{x},\hat{x}}\hat{T}(\hat{x},\hat{t})=\zeta  \hat{J}(\hat{t})^2
\end{equation}
with 
%\begin{subequations}
\begin{equation}
\hat{T}(0,\hat{t})=\hat{T}^{(0)}_\text{h} +\hat{T}^{(1)}_\text{h} \phi(\hat{t},\hat{\omega};\beta), \qquad %\label{eq:DBC1a}\\
%\end{equation}
%\begin{equation}
\hat{T}(1, \hat{t}) = \hat{T}_\text{c} %\label{eq:DBC2a}
%\end{equation}
%\begin{equation}
\end{equation}
%\end{subequations}
%Suppressing all hats, 
with initial condition $\hat{T}(\hat{x},0)=(1-\x)+\x \T_\text{c}$.
Integration returns
\begin{equation}
\begin{array}{lll}
\T(\x,\ttt) &=& (1-\x)(1+\T^{(1)}_\text{h} \phi(\ttt,\hat{\omega};\beta))\\
&\phantom{=}&+ \x \T_\text{c}+\ds\sum_{n=1}^{\infty }{D_n(\ttt)\sin (n \pi \x)} 
\end{array}
\end{equation}
with 
\begin{equation}
D_n(\ttt,\hat{\omega};\zeta)=-\frac{2}{n\pi} \int_0^{\hat{t}} e^{-n^2\pi ^2 (\ttt-\tau )} \left(\T^{(1)}_\text{h} \phi'(\tau,\hat{\omega};\beta)+\zeta  \hat{J}(\tau )^2 \gamma _n\right) \, \text{d}\tau
\label{eq:dn}
\end{equation}
and $\gamma_n=-1+(-1)^n$.

The \emph{instantaneous} balance of heat fluxes entering and leaving the medium accounts then to
\begin{equation}
\begin{array}{lll}
\Delta \hat{J}_\text{q}(\ttt) &=&
\zeta \hat{J}(\ttt)\ds\left(1+\T^{(1)}_\text{h} \phi(\ttt,\hat{\omega};\beta)-\T_\text{c}\right)\\
&\phantom{=}& +\ds\sum_{n=1}^{\infty }n \pi  \gamma_n D_n(\ttt,\hat{\omega};\zeta)
\end{array}
\label{eq:deltajq}
\end{equation}
In turn, the current density depends on the temperature drop across the leg as $\hat{J}(\ttt)=\left(1+ \T^{(1)}_\text{h} \phi(\ttt,\hat{\omega};\beta)-\T_\text{c}\right)/(m+1)$, where $m$ is the ratio between the load electric resistance $R_\text{L}$ and the leg electric resistance $R_\text{TE}$. Thus, replacing it into Eqs.\ \eqref{eq:dn} and \eqref{eq:deltajq} provides a closed form of $\Delta \hat{J}_\text{q}(\ttt)$. 
Notably, the instantaneous heat balance is not zero even for $\zeta=0$, as heat is temporarily stored in the medium. However, its averaged value over a cycle is zero for a non-thermoelectric material. Instead, when $\zeta\ne 0$, energy conservation returns the cycle-averaged power output density 
\begin{equation}
\langle \hat{J}_\text{w}(\hat{t}) \rangle =
\langle \Delta \hat{J}_\text{q} (\ttt) \rangle =\frac{\hat{\omega}}{2\pi}
\int_0^{2\pi/\hat{\omega}} \Delta \hat{J}_\text{q} (\ttt) \text{d}\ttt
\end{equation}
Integration returns
\begin{equation}
\langle \hat{J}_\text{w} (\hat{t})\rangle =
\frac{\zeta  m}{(m+1)^2} \left[\left(1-\T_\text{c}\right)^2+(1-2 \beta ) \T_\text{h}^{(1)\; 2}\right]
%\frac{\zeta  \left(\left(1-T_\text{c}\right){}^2+(1-2 \beta ) T^{(1)\; 2}_\text{h}\right)}{m+1}
%-\frac{\zeta  \left(\left(1-T_\text{c}\right){}^2+(1-2 \beta ) T^{(1)\; 2}_\text{h}\right)}{(m+1)^2}
\end{equation}
Maximum output power density is easily computed by differentiating over $m$ and setting the derivative to zero. One finds $m=1$, as in the stationary case. Thus, maximum output power density reads
\begin{equation}
\langle \hat{J}_\text{w} (\hat{t})\rangle_\text{MP} =
\frac{\zeta}{4}  \left[\left(1-\T_\text{c}\right)^2+(1-2 \beta ) \T^{(1)\; 2}_\text{h}\right]
\end{equation}
or, reverting to dimensioned quantities,
\begin{equation}
\langle J_\text{w} (t)\rangle_\text{MP} =
\frac{\sigma_T \alpha^2}{4\ell} \Delta T^{(0)\,2}  \left[1+(1-2 \beta) \theta^2\right]
%\frac{\sigma_T \alpha^2}{4\ell}  \left[\left(T_\text{h}^{(0)}-T_\text{c}\right)^2+(1-2 \beta ) T^{(1)\; 2}_\text{h}\right]
\end{equation}
where $\Delta T^{(0)}=T_\text{h}^{(0)}-T_\text{c}$ and $\theta=T_\text{h}^{(1)}/\Delta T^{(0)}$.
Consistently, by setting $T_\text{h}^{(1)}$ to zero one recovers the stationary power output density $J_\text{w,ss} = \sigma_T \alpha^2/(4\ell)\Delta T^{(0)\,2}$ while for $\beta=1/4$ we obtain the average power output density already reported for a sinusoidal temperature modulation \cite{Narducci2025}, $\langle J_\text{w} \rangle_\text{MP,sine} = \sigma_T \alpha^2/(4\ell) \Delta T^{(0)\,2}  \left[1+\frac{1}{2} \theta^2\right]$.
Output power density remains independent of the modulation frequency for any $\beta$ value.
Efficiency at maximum power averaged over a cycle $\eta_\text{MP}=\langle\hat{J}_\text{w}\rangle/\langle\hat{J}_\text{q}(0)\rangle$ reads instead
\begin{equation}
\eta_\text{MP}
=\frac{\eta_\text{C}^{(0)}}{2}\frac{1+(1-2\beta)\theta^2}{1+2/\zeta-\frac{1}{4}\eta_\text{C}^{(0)}\left[1-3(1-2\beta)\theta^2\right]}
\end{equation}
where $\eta_\text{C}^{(0)}=1-\hat{T}_\text{c}=\Delta T^{(0)}/T_\text{h}^{(0)}$ is Carnot's stationary efficiency.

\begin{figure}
\includegraphics[width=.9\columnwidth]{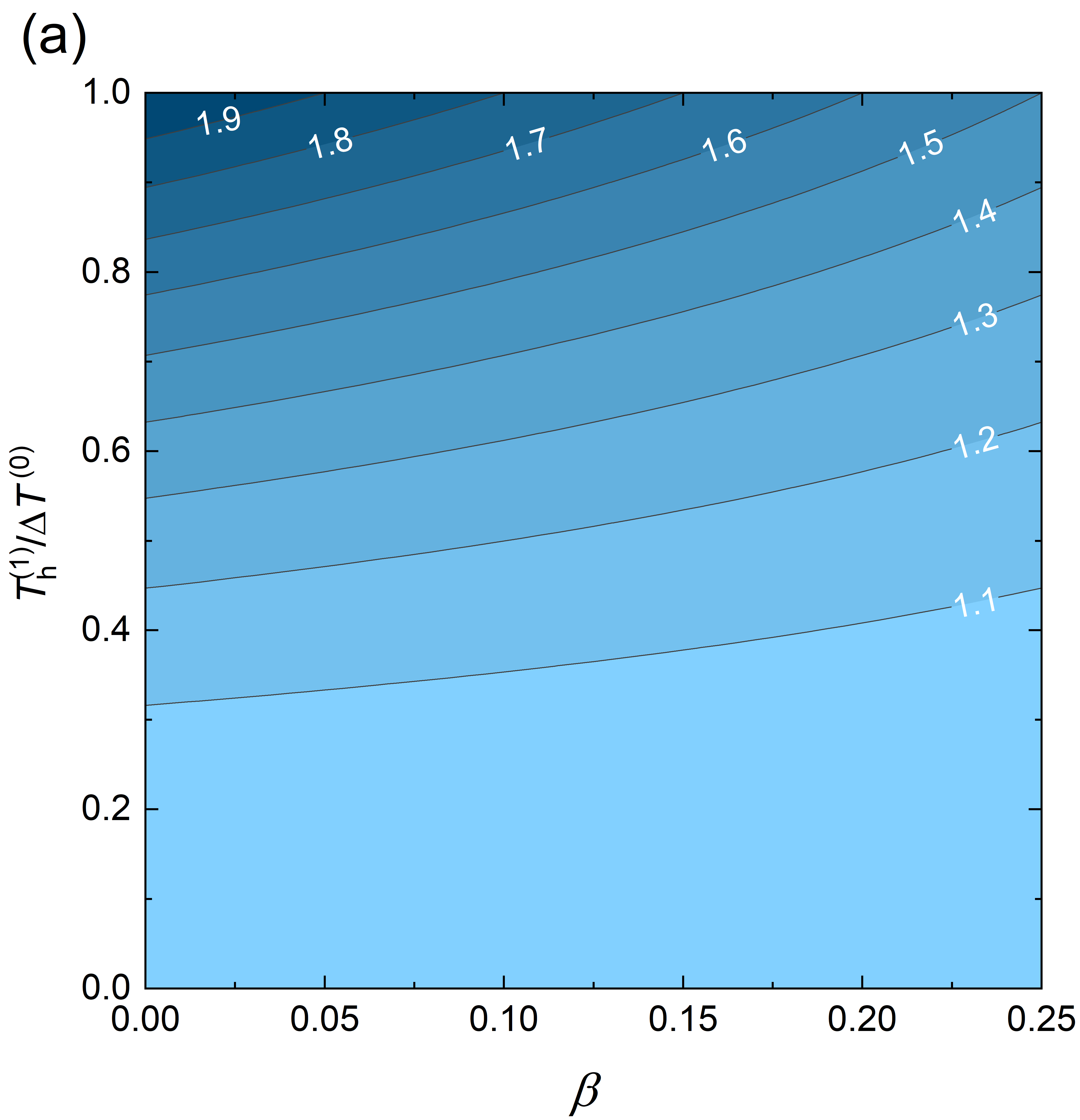}
\includegraphics[width=.9\columnwidth]{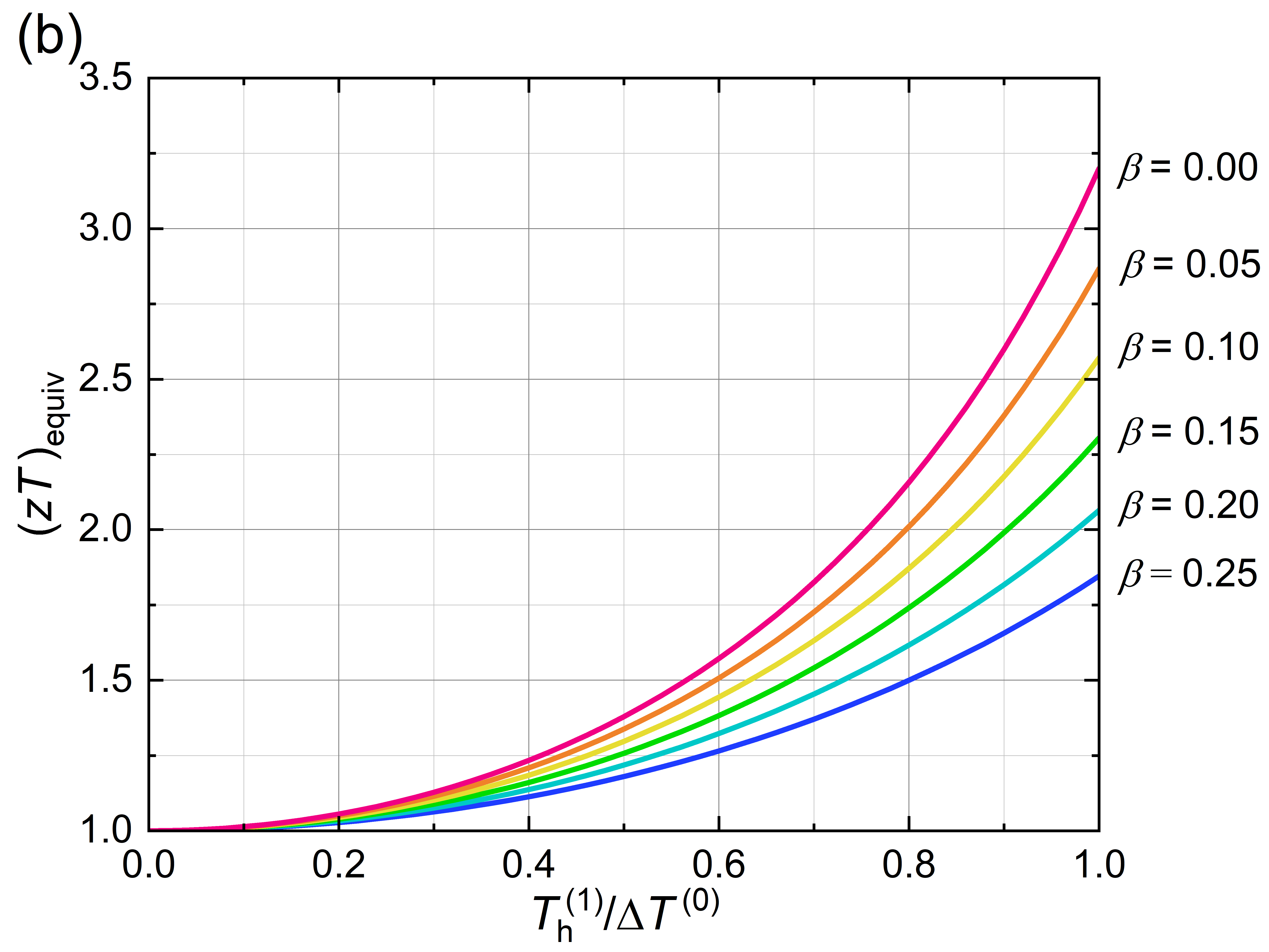}
\caption{
(a) Ratio between the average output power density $\langle J_\text{w} \rangle_\text{MP}$ and the stationary output power density $J_\text{w,ss}$ and (b) equivalent figure of merit $(zT)_\text{equiv}$ of a TEG with $zT_\text{h}^{(0)}=1$, $T_\text{h}^{(0)}=400$ K as a function of the modulation amplitude $T^{(1)}_\text{h}$ normalized to the average temperature difference $\Delta T^{(0)}$ and of the smoothing factor $\beta$.}
%\label{fig:zT}
\label{fig:fig2}
\end{figure}

Figure \ref{fig:fig2}(a) shows the ratio between $\langle J_\text{w} \rangle_\text{MP}$ and  $J_\text{w,ss}$ as a function of $T_\text{h}^{(1)}$ and $\beta$. It displays how the maximum power output is attained with the largest modulation amplitude $T_\text{h}^{(1)}$ and for $\beta\to 0$, namely in the square wave limit, as anticipated. A doubled power output is attained. We define an equivalent thermoelectric figure of merit $(zT)_\text{equiv}$ as the $zT$ value that a TEG converting heat into electric power under conventional stationary conditions should have to output the average electric power it generates when a time-modulated temperature difference $T_\text{h}^{(0)}+ T^{(1)}_\text{h} \phi(t,\omega;\beta)-T_\text{c}$ is applied. For $zT=1$, $(zT)_\text{equiv}$ is increased up to the remarkable value of 3.2 (Fig.\ \ref{fig:fig2}(b)). 
A fully comparable trend is reported by the EMP. Doubled output power density converts into a maximum EMP enhancement of 88\% for $\beta\to 0$, since time-modulation slightly increases the input heat flux (by $<$5\%). 

\section{Dynamic Efficiency Beyond the Constant-Property Approximation}

As well known, CPA holds exactly true only in the $\Delta T\to 0$ limit. Thus, it may be then interesting to re-analyze the effect of TEG dynamic operation beyond the CPA. To this aim, we chose the exemplary case of polycrystalline Bi$_{0.5}$Sb$_{1.5}$Te$_3$, a $p$-type thermoelectric material largely used in commercial devices.
To account for the temperature dependence of $\sigma$, $\alpha$, $\kappa_\text{oc}$, and $c_V$ (Table \ref{tab:1}), since their values depend on position and time through $T(x,t)$, we dealt with the medium as an infinite cascade of thermoelectric elements. Integration of Domenicali's equation in its dimensioned form [Eq.\ \eqref{eq:tdDE}] is carried out by keeping $J(t)$ as a parameter, obtaining $T(x,t;J(t))$.  Feeding it into
\begin{equation}
J(t)=
\ds\frac{\alpha(T(\ell,t))T(\ell,t)-\alpha(T(0,t))T(0,t)}{2\ell\bar{\rho}(t)}
\label{eq:JnonCPA}
\end{equation}
(where $\ell$ is the leg length) one obtains an implicit equation in $J(t)$ whose solution returns the current density at time $t$. In the previous equation, the effective resistivity $\bar{\rho}(t)$ is obtained as follows.
Let us consider the leg as a segmented series of $N$ thermoelectric elements, all traversed by the same $J(t)$. At each time, the contribution of the $i$-th element to the total thermovoltage $\Delta V_\text{tot}(t)$ is 
$\Delta V_i(t) = 2\ell\rho(T(x_i,t))J(t)$,
with 
\begin{equation}
\Delta V_i(t)=\alpha(T(x_i,t),t)T(x_i,t)-\alpha(T(x_{i+1},t),t)T(x_{i+1},t)
\end{equation} 
Since 
\begin{equation}
\Delta V_\text{tot}(t)=\sum_i \Delta V_i(t)= \alpha(\ell,t))T(\ell,t)-\alpha(T(0,t))T(0,t)
\end{equation}
then
\begin{equation}
\sum_i\ell\rho_i(t)=\sum_i \Delta V_i(t)/J(t)=\Delta V_\text{tot}(t)/J(t)
\end{equation} 
Comparing it to Eq.\ \eqref{eq:JnonCPA} one finds that $\bar{\rho}(t)=\sum_i\rho_i(t)$. For $N\to\infty$ this leads to
\begin{equation}
\bar{\rho}(t)=\ell^{-1}\int_0^\ell\sigma(T(x,t))^{-1}\text{d}x
\end{equation}
Computation of instantaneous heat fluxes entering and leaving the medium follow as
\begin{equation}
J_\text{q}(x)=-\kappa(T(x,t))\partial_x T(x,t)+\alpha(T(x,t))T(x,t)J(t)
\end{equation}
for $x=0, \ell$. This lets obtain the instantaneous power output density as $\Delta J_\text{q}=J_\text{w}(t)=J_\text{q}(0)-J_\text{q}(\ell)$, which is then averaged over a cycle to obtain $\langle J_\text{w}(t) \rangle_\text{MP}$, to 
compare to the steady-state power output density.

\begin{figure}
\includegraphics[width=.4\textwidth]{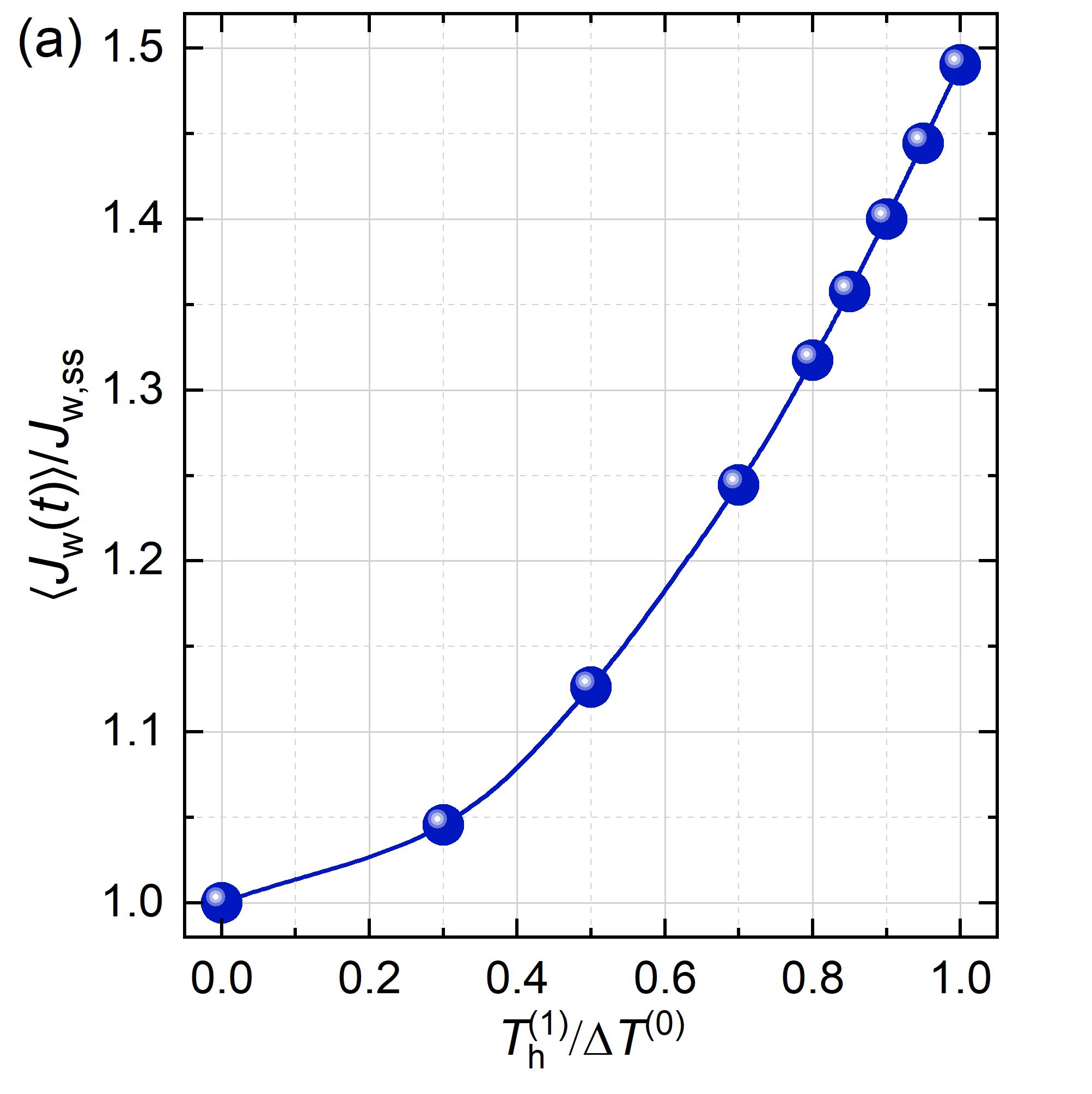}\\
\hspace{-.9cm}\includegraphics[width=.4\textwidth]{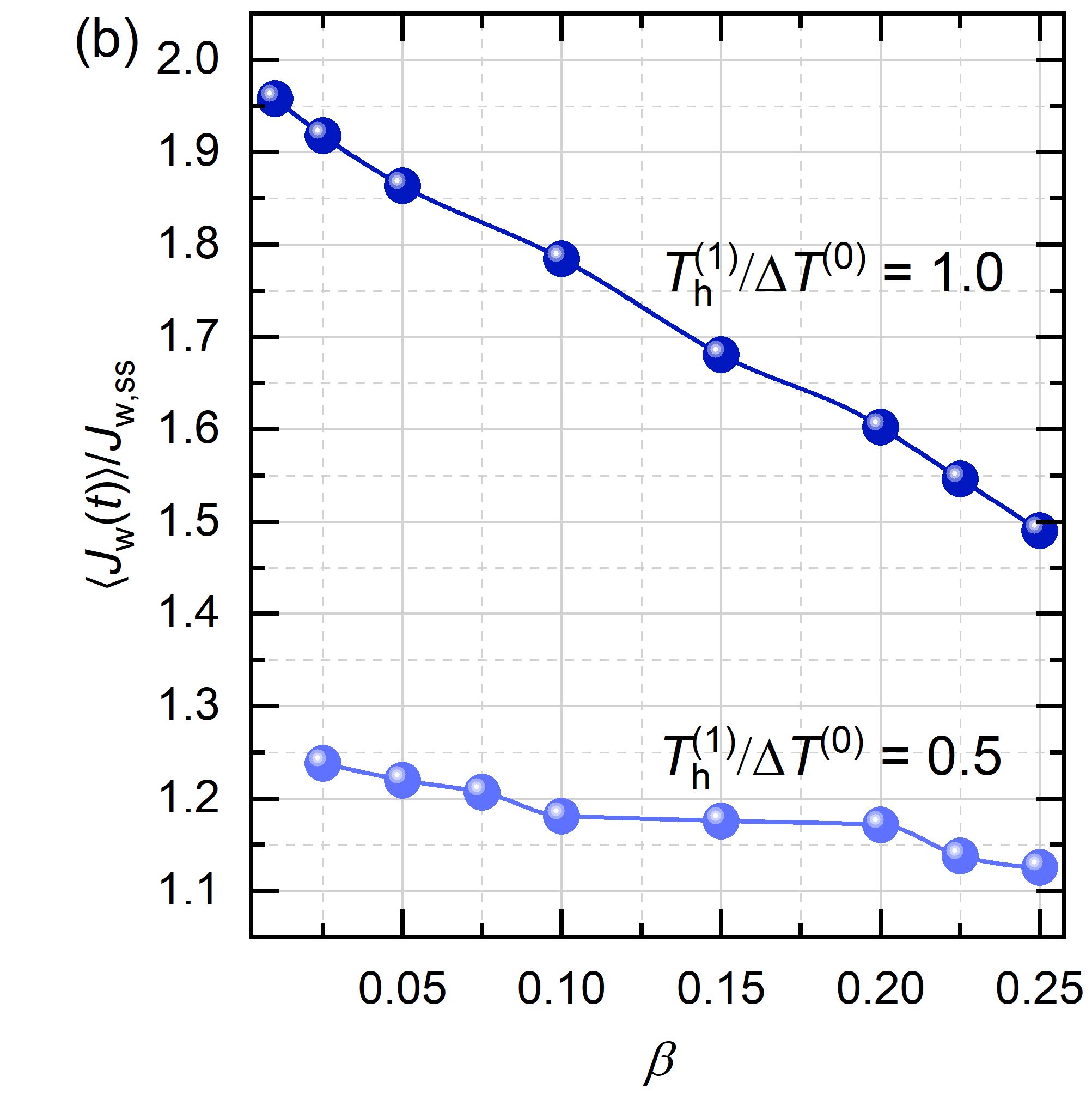}
\caption{Non-CPA values of the power output density averaged over the cycle and normalized to its stationary value for Bi$_{0.5}$Sb$_{1.5}$Te$_3$ (a) as a function of the modulation amplitude for $\beta=1/4$ (sinusoidal modulation), $\omega=1$ Hz; and (b) as a function of $\beta$, still for $\omega=1$ Hz. In both plots we set $T_\text{h}^{(0)}=400$ K, $T_\text{c}=300$ K, $\ell =3$  mm and $m=1$.}
\label{fig:fig3}
\end{figure}

\begin{table}
\caption{Interpolating formulas of $\sigma(T)$, $\alpha(T)$, and $\kappa_\text{oc}(T)$ of Bi$_{0.5}$Sb$_{1.5}$Te$_3$ 
\cite{data}. Temperatures in K.}
\label{tab:1}
\begin{ruledtabular}
\begin{tabular}{lll}
Quantity & Units & Formula\\
\hline
$\sigma$& $\Omega^{-1}$m$^{-1}$ & $(2.87619\times 10^4) T^{0.2}$\\
$\alpha$& V/K & $2.40\times 10^{-4} \left(1. -3\times 10^{-6} (T-380.)^2\right)$\\
$\kappa_\text{oc}$& W/mK & $1.5\times 10^{-3} (T-300.)+1.6$\\
$c_V$ & J m$^{-3}$ K$^{-1}$ & $1.155\times 10^3 T + 7.315\times 10^5$
\end{tabular}
\end{ruledtabular}
\end{table}

%\begin{figure*}
%\includegraphics[width=.8\textwidth]{fig4.jpg}
%\caption{(a) Non-CPA power output density averaged over the cycle and normalized to its stationary value for Bi$_{0.5}$Sb$_{1.5}$Te$_3$ as a function of $\omega$; (b) Change of leg resistivity (averaged over its length and normalized to its stationary value) vs.\ time over the cycle for $\omega=$ 0.1, 1.0, and 10 Hz with $\beta=0.01$. Note how the resistivity is smaller at low frequencies in the half-cycle where $T(t)>T_\text{h}^{(0)}$ (top plot). Data computed for $T_\text{h}^{(0)}=400$ K,$T_\text{h}^{(1)}=100$ K, $T_\text{c}=300$ K, $\ell =3$  mm and $m=1$.}
%\label{fig:fig4}
%\end{figure*}

Figure \ref{fig:fig3}(a) displays  $\langle J_\text{w}(t) \rangle_\text{MP}/ J_\text{w, ss}$ normalized to the stationary value of the power output density as function of the temperature modulation amplitude. Its enhancement is confirmed and is minimally altered by the temperature dependence of the phenomenological coefficients.
We also observe an almost linear dependence of $\langle J_\text{w}(t) \rangle_\text{MP}/ J_\text{w, ss}$ on $\beta$, with the power output doubling with respect to the stationary case when $\beta\to 0$, fully mirroring the CPA case (Fig.\ \ref{fig:fig3}(b)). 
As in the CPA, no frequency dependence is observed.

\section{Concluding remarks}
In summary, we have shown that, within the CPA, time-modulation of the temperature difference applied to a thermoelectric medium causes an increase of the average output power density (efficiency at maximum power) of up to 100 \% (88 \%) compared to the very same system operated under stationary conditions -- an improvement equivalent to increase its figure of merit from 1 to 3.2. 
Concerning real materials (beyond the CPA), since in most thermoelectric materials an increase of temperature causes an increase of $z(T)T$, the analysis strengthens the conclusion that a large increase of the power output may be achieved under dynamic boundary conditions, even exceeding the already large value of the effective figure of merit computed in the CPA limit.
Note that such outstanding increase applies to any thermoelectric material and does not imply any new device design, and is therefore immediately applicable to any commercial thermoelectric generator.

We finally mention that it was shown that dynamic thermoelectric conversion does not require innately time-modulated heat sinks, since modulation may be imparted by properly engineering the heat flux the source discharges toward the environment, yet with no additional losses of thermal power \cite{Narducci2025}. Thus, the present results should be considered as a novel way to  overcome the often-complained low efficiency of thermoelectric generators.


% If in two-column mode, this environment will change to single-column
% format so that long equations can be displayed. Use
%\begin{widetext}
%\end{widetext}


%\bibliography{c:/dox/tex/bibtex/Seebeckbib} %,biblio
\bibliography{Seebeckbib} %,biblio

\end{document}